\documentstyle[12pt]{article}
\def\bq{\begin{equation}}       \def\eq{\end{equation}}
\def\bqn{\begin{eqnarray}}      \def\eqn{\end{eqnarray}}
\def\n{\nu}                     \def\m{\mu}
\def\f{\phi}                    \def\t{\theta}
\def\G{\Gamma}                  \def\g{\gamma}

\begin{document}

\thispagestyle{empty}
\begin{flushright}
CINVESTAV-FIS-06/96
\end{flushright}
\vspace{2cm}
\begin{center}
{\small\bf The Number of Neutrinos and the Left-Right Symmetric Model}
\end{center}

\vspace{1cm}

\begin{flushleft}
{R. Huerta$^1$, M. Maya$^2$ and O. G. Miranda$^3$ \\
$^1$Departamento de F\'{\i}sica Aplicada Centro de Investigaci\'{o}n y de 
Estudios Avanzados del IPN Unidad M\'{e}rida, 
A. P. 73 Cordemex 97310, M\'{e}rida, Yucat\'{a}n, M\'{e}xico, 
(e-mail: rhuerta@kin.cieamer.conacyt.mx). \\
$^2$Facultad de Ciencias F\'{\i}sico Matem\'{a}ticas 
Universidad Aut\'{o}noma de Puebla\\A. P. 1364, Puebla, 7200, M\'{e}xico. \\
$^3$Depto. de F\'{\i}sica, Centro de Investigaci\'{o}n y de 
Estudios Avanzados del IPN A. P. 14-740, M\'{e}xico 07000, D. F.,
M\'{e}xico \\
(e-mail: omr@fis.cinvestav.mx). }
\end{flushleft}

\vspace{1cm}

\begin{abstract}
{The meaning of $N_\n$ in the Left-Right Symmetric Model (LRSM) is discussed. 
In its original definition $N_\n$ is the number  of neutrinos or 
generations of leptons, so, in the Standard Model its value is 
three. However the determination of $N_\n$ in experiments and in 
astrophysical observations is subject to different theoretical 
interpretations. We present arguments that gives $N_\n$ as a 
parameter of the LRSM. Using an 
experimental value for the rate $\Gamma_{\rm inv}/\Gamma_{l\bar{l}}$ 
we calculate the bound $2.90\leq N_{\n_{LR}}\leq 3.04$ (90 \% C. L.), where
$N_{\n_{LR}}$ is a function of the left-right mixing angle $\f$.
This range is less restrictive 
than the one of the standard model.}
\end{abstract}

\vspace{1cm}
\begin{center}
{\it to be published in {\bf Z. Phys. C}}
\end{center}
\newpage

\baselineskip=25pt

\section{Introduction}
Many models have been proposed in the literature that give the
standard model (SM) [1] in the low energy limit. One of these models is the
left-right symmetric model (LRSM) [2], with a $SU_L(2)\times SU_R(2)\times
U(1)$ gauge group. In this class of models the gauge symmetry is
spontaneously broken giving different masses to the left and 
right-handed
gauge bosons. In the low-energy limit the LRSM becomes the SM and small
deviations could be observed making high precision experiments. 
In order to know how many light neutrinos have the SM we need to know 
the value 
of the  partial decay width $\Gamma(Z\rightarrow\n\bar{\n})$. However, 
this process is such that no final states are observed. Hence, the only 
way to get information about this partial decay is from the invisible 
width $\Gamma_{\rm inv}=\Gamma_Z-(\Gamma_{\rm had}+\Gamma_{ee}+
\Gamma_{\mu\mu} + \Gamma_{\tau\tau})$. The number $N_\n$ is 
defined as a combination of the experimental and the theoretical 
magnitudes of 
$\Gamma_{\rm inv}$, $\Gamma_{l\bar{l}}$ and 
$\Gamma_{\n\bar{\n}}$ and it is then identified with the number of light 
neutrinos. 

Recent 
experimental analysis  on $N_\n$, in the framework of the SM, gives as a 
result $N_\n =2.983\pm 0.034$.
In this paper we make the calculation of $N_\n$ in the LRSM. We found 
that in this case the quantity defined as $N_\n$ is not a constant but 
depends on the mixing angle 
$\f$ and therefore is not necessarily an integer number. 
We give an estimation of the value $N_\n$ in this model using 
experimental data and also from previous constraints on the $\f$ angle.

In Sec. 2 we describe the model, whereas in Sec. 3 we discuss the meaning of
$N_\n$ in the LRSM and give the results of the calculation and also our
conclusions.
%$$$$$$$$$$$$$$$$$$$$  The LR Model  $$$$$$$$$$$$$$$$$$$$$$$$$$$$$$$$$$
\section{The Left-Right Symmetric Lagrangian}
In the LRSM the Lagrangian that describes the interaction between
the leptons and the neutral gauge bosons is
%%%%%%%%% Ec (1) %%%%%%%%%%%%%%%%%%%%%%%%%%%%%%%%%%%%%%%%%%
\bq  {\cal L}_N=gJ_L^3W_L^3+gJ_R^3W_R^3+\frac{1}{2}g'J_YB  \eq
where $W_R^3$ is the neutral gauge boson of the $SU_R(2)$ sector of the model
that couples with the right-handed current $J_R^3$ of leptons. 
We are interested in the LRSM with a bidoublet and two doublets in the 
Higss sector; this Higss content necessarily implies Dirac Neutrinos. 
After the spontaneous breakdown of the symmetry, the gauge bosons
become mass eigenstates and therefore linear combinations of $W_L^3$, $W_R^3$
and $B$. The relation between the mass eigenstates and the interaction
eigenstates is given through a mixing matrix. This mixing matrix depends
on two angles: $\t_W$, the SM electroweak parameter, and $\f$, that mixes the
left and right handed gauge bosons, after imposing the condition that
the electromagnetic current has to couple to the photon. The mixing
can be realized in two steps [4]:
%%%%%%%%% Ec (2) %%%%%%%%%%%%%%%%%%%%%%%%%%%%%%%%%%%%%%%%%%
\bqn\left(\begin{array}{c}Z_1\\Z_2\\A\end{array}\right)=
    \left(\begin{array}{ccc}c_\f&-s_\f&0\\s_\f&c_\f&0\\0&0&1
\end{array}\right)
\left(\begin{array}{ccc}c_W&-s_Wt_W&-t_Wr_W\\0&r_W/c_W&-t_W\\s_W&s_W&r_W
\end{array}\right)\left(\begin{array}{c}W^3_L\\W^3_R\\B\end{array}\right)\eqn
with $s_\f=\sin\f$, $c_\f=\cos\f$, $s_W=\sin\t_W$, $c_W=\cos\t_W$ and
$r_W=\sqrt{\cos2\t_W}$. 
In terms of the mass eigenstates the general Lagrangian responsible for the
neutral current interactions is written  as
%%%%%%%%% Ec (3) %%%%%%%%%%%%%%%%%%%%%%%%%%%%%%%%%%%%%%%%%%
\bq  {\cal L}_N=eJ_{em}A-\frac{e}{c_\t}(a_1J_L^Z+b_1J_R^Z)Z_1
               +\frac{e}{c_\t}(a_2J_L^Z+b_2J_R^Z)Z_2  \eq
where
%%%%%%%%% Ec (4) %%%%%%%%%%%%%%%%%%%%%%%%%%%%%%%%%%%%%%%%%%
\bqn  a_1&=&\frac{s_Ws_\f}{r_W}-\frac{c_\f}{s_W}\nonumber\\
      b_1&=&\frac{c_W^2s_\f}{s_Wr_W}             \nonumber\\
      a_2&=&\frac{s_Wc_\f}{r_W}+\frac{s_\f}{s_W}  \nonumber\\
      b_2&=&\frac{c_W^2s_\f}{s_Wr_W}                    \eqn
In the limit $\f\rightarrow 0$ and $M_{Z_2}\rightarrow\infty$ we
get the SM Lagrangian. In Eq. (3) we have all the information we need to
calculate the process $Z_1\rightarrow f\bar{f}$.
%$$$$$$$$$$$$$$$$$$$$  N nu como Funcisn de phi  $$$$$$$$$$$$$$$$$$$$$$$$
\section{$N_\n$ as a function of $\f$}
If we are working at the $Z$ peak, the amplitude for the decay
$Z_1\rightarrow l\bar{l}$ comes from the second term of (3) and is given
by [4]
%%%%%%%%% Ec (5) %%%%%%%%%%%%%%%%%%%%%%%%%%%%%%%%%%%%%%%%%%
\bq M=\frac{g}{c_W}
      \left[\bar{u}\g^\m\frac{1}{2}(g_{V_{LR}}-g_{A_{LR}}\g_5)v\right]
      \epsilon_\m^\lambda   \eq
where $\epsilon_\m^\lambda$ is the $Z_1$ boson polarization, $u$ $(v)$
is the lepton (antilepton) spinor and
%%%%%%%%% Ec (6) %%%%%%%%%%%%%%%%%%%%%%%%%%%%%%%%%%%%%%%%%%
\bq  g_{V_{LR}}=\left[c_\f-\frac{s_W^2}{r_W}s_\f\right]\bar{g}_V
     -\frac{c_W^2}{r_W}s_\f g_{V_R},  \eq
%%%%%%%%% Ec (7) %%%%%%%%%%%%%%%%%%%%%%%%%%%%%%%%%%%%%%%%%%
\bq  g_{A_{LR}}=\left[c_\f-\frac{s_W^2}{r_W}s_\f\right]\bar{g}_A
     +\frac{c_W^2}{r_W}s_\f g_{A_R}   \eq
Here $\bar{g}_V$ ($\bar{g}_A$) is the vector (axial-vector) coupling
constant for charged leptons with radiative corrections that comes
from the left-handed sector of the model:
%%%%%%%%% Ec (8) %%%%%%%%%%%%%%%%%%%%%%%%%%%%%%%%%%%%%%%%%%
\bqn
\bar{g}_V&=&\sqrt{\rho_f}\left(-\frac{1}{2}+2\kappa_f\sin^2\theta_W\right)
      \nonumber\\  \bar{g}_A&=&\sqrt{\rho_f}\left(-\frac{1}{2}\right)   \eqn
while $g_{V_R}$ ($g_{A_R}$) is the same coupling constant as before but  
without 
radiative corrections and has its origin in the right-handed sector.\\
For the decay of $Z_1$ in neutrinos we also have corrections for the
coupling constants that comes only from the LRSM:
%%%%%%%%% Ec (9) %%%%%%%%%%%%%%%%%%%%%%%%%%%%%%%%%%%%%%%%%%
\bq  g_{V_{LR}}^\n=\left(c_\f-\frac{s_\f}{r_W}\right)g^\n   \eq
and
%%%%%%%%% Ec (10) %%%%%%%%%%%%%%%%%%%%%%%%%%%%%%%%%%%%%%%%%%
\bq  g_{A_{LR}}^\n=\left(c_\f+s_\f r_W\right)g^\n  \eq
where $g^\n=\frac12$.\\
In the framework of the LRSM model, the width for the process
$Z\rightarrow l\bar{l}$ is written as
%%%%%%%%% Ec (11) %%%%%%%%%%%%%%%%%%%%%%%%%%%%%%%%%%%%%%%%%%
\bq  \Gamma(Z\rightarrow l\bar{l})=\frac{G_FM_Z^3}{6\pi\sqrt{2}}
     \left[g_{V_{LR}}^2+g_{A_{LR}}^2\right]   \eq
In Eq. (11) we have neglected the mass of the leptons.
In the case of the decay $Z\rightarrow\n\bar{\n}$ the expression (11) becomes
%%%%%%%%% Ec (12) %%%%%%%%%%%%%%%%%%%%%%%%%%%%%%%%%%%%%%%%%
\bq  \Gamma(Z\rightarrow\n\bar{\n})=\frac{G_FM_Z^3}{6\pi\sqrt{2}}
     (g^\n)^2\left[\left(c_\f-\frac{s_\f}{r_W}\right)^2
     +\left(c_\f+r_Ws_\f\right)^2\right]   \eq
In the limit when $\f$ goes to zero we obtain the SM
tree level partial decay width:
%%%%%%%%% Ec (13) %%%%%%%%%%%%%%%%%%%%%%%%%%%%%%%%%%%%%%%%%
\bq  \Gamma_0(Z\rightarrow\n\bar{\n})=\frac{G_FM_Z^3}{12\pi\sqrt{2}}    \eq
The partial widths (11) and (12) are applicable to all charged leptons and
all neutrinos respectively. In order to use the expressions (11) and (12) for
comparing with the experimental result for the number of light neutrinos
$N_\n$ we recall the experimental definition for $N_\n$ in a SM analysis [7],
%%%%%%%%% Ec (14) %%%%%%%%%%%%%%%%%%%%%%%%%%%%%%%%%%%%%%%%%
\bq  N_\n=R_{\rm exp}
          \left(\frac{\G_{l\bar{l}}}{\G_{\n\bar{\n}}}\right)_{SM}. \eq
Here, the quantity in parenthesis is the standard model prediction and the
$R_{\rm exp}$ factor is the experimental value of the ratio between the widths
$\G_{\rm inv}$ and $\G_{l\bar{l}}$ [5],
%%%%%%%%% Ec (15) %%%%%%%%%%%%%%%%%%%%%%%%%%%%%%%%%%%%%%%%%
\bq  R_{\rm exp}=\frac{\G_{\rm inv}}{\G_{l\bar{l}}}=5.942\pm 0.067.  \eq
The definition (14) for $N_\n$ replaces the expression
$N_\n=\G_{\rm inv}/\G_{\n\bar{\n}}$, since (14) reduces the influence of
the top quark mass.\\
If we want to get information about what is the meaning of $N_\n$ in the LRSM
model we should define the corresponding expression,
%%%%%%%%% Ec (16) %%%%%%%%%%%%%%%%%%%%%%%%%%%%%%%%%%%%%%%%%
\bq  N_{\n_{LR}}=R_{\rm exp}
     \left(\frac{\G_{l\bar{l}}}{\G_{\n\bar{\n}}}\right)_{LR}\eq
This new expression will be a function of $\f$, so in this case the 
quantity 
defined as the number of light neutrinos is not a constant and not
necessarily an integer. Also $N_{\n_{LR}}$ in formula (16) is 
independent from the $Z_2$ mass and therefore depends on only 
one parameter of the LRSM. 
Experimental values for 
$\Gamma_{\rm inv}$ and for $\G_{l\bar{l}}$ are reported in 
literature which in our case, can give a 
bound for the angle $\f$. However, we can look to these experimental 
numbers in
another way. The partial widths $\Gamma_{\rm inv}=499.9\pm2.5$ and
$\G_{l\bar{l}}=83.93\pm0.14$ were reported recently [8], but we
use here the value (15) for the $R_{\rm exp}$ rate of Ref. [5].
All these measurements are independent of any model and can be fitted with
the LRSM parameter $N_{\n_{LR}}$ in terms of $\f$. We can plot the expression
(16) to see the general behavior of the
$N_{\n_{LR}}(\f)$ function. The Fig. (1) is this plot.
We can observe that for some values of $\f$, around 0.6 rad, $N_{\n_{LR}}$
can be as high as 5.9, and for values of $\f$ around
$-0.9$ rad, $N_{\n_{LR}}$ is as low as 0. This indicates a strong dependence on
$\f$ for leptonic decays of $Z$. Therefore, according to the above discussion,
if we consider $N_{\n_{LR}}$ as the number of neutrinos, the
restriction on the number of generations can be "softened" if we consider
a LRSM model.\\
However $\f$ is severely bounded, so it is $N_{\n_{LR}}$.
In Fig. (2) we show the allowed region for $N_{\n_{LR}}$ with 90\% C. L.
The allowed region is the inclined band that is a result of both factors
in Eq. (16): $(\G_{l\bar{l}}/\G_{\n\bar{\n}})_{LR}$ gives the
inclination and $R_{\rm exp}$ gives the broading. The analysis was done using
the experimental value (15) for $R_{\rm exp}$ reported by
ALEPH Collaboration [5] with a 90\% C. L. The region for $\f$ is given from 
a previous constraint on this angle [4]: $-0.009\leq\f\leq0.004$,
obtained from the LEP experimental value of $g_A$ [7].
In the same figure we show the SM ($\f=0$) result at 90\% C. L. with the
solid horizontal lines. As we can see, 
the  allowed region in the LR model (dotted line)
for $N_{\n_{LR}}$ is wider that the one for the SM, and is 
given by: 
%%%%%%%%% Ec (17) %%%%%%%%%%%%%%%%%%%%%%%%%%%%%%%%%%%%%%%%%
\bq    2.90\leq N_{\n_{LR}}\leq 3.04.   \eq
Although the allowed region is wider, the fact that the
mixing angle has been extremely restricted forces $N_{\n_{LR}}$ to be near 
3, excluding the possibility to get a number such as 2.6 which has
been reported as an upper bound from big bang nucleosynthesis [9]. In 
order to get this value we need $\f =-0.063$, which is in contradiction 
with the present restrictions on $\f$.

Finally, and just for completeness, we can reverse the arguments, that 
is, we fix the number of neutrinos in the LRSM to be three
then the theoretical expression
for $R$ will be given by 
\bq
R=\frac{3\Gamma_{\n\bar{\n}}}{\Gamma{l\bar{l}}}.
\eq
If we make a plot of this quantity in terms of $\f$ we have the curve 
shown in Fig.(3). The horizontal lines give the experimental region at 90 \% 
C. L. As we can see from the figure the constraint for the $\f$ angle is: 
$-0.006\leq\f\leq 0.011$. This restriction is in close agreement with the 
constraint previously obtained in Ref. [4].\\
As a conclusion we can say that in the left-right symmetric model we can 
obtain
from experimental results a value for $N_\n$ different from 3 (not necessarily
an integer number). In particular for the LRSM with two doublets, and hence
with Dirac neutrinos, $N_{\n_{LR}}$ is in the neighborhood of three, However, 
if 
new precision experiments 
find small deviations from 3, this model could explain
very well these deviations with a small value of $\f$ . 

{\bf Acknowledgments} \newline
This work was supported in part by CONACYT (M\'{e}xico).

\newpage
\noindent{\Large\bf Figure Captions}
\vspace{1cm}

\noindent Fig. 1 $N_{\n_{LR}}$ as a function of $\f$ (rad). 

\vspace{0.5cm}

\noindent Fig. 2 Allowed region for $N_{\n_{LR}}$ from the experimental value 
$R_{\rm exp}$ and from a previous constraint for $\f$. The dashed line shows 
the SM allowed region for $N_\n$ at 90\% C. L. while the dotted line shows 
the same result for the LRSM.

\vspace{0.5cm}

\noindent Fig. 3 The curve shows the shape for $R$ in the LRSM. The dashed line shows the experimental region at 90 \% C. L.

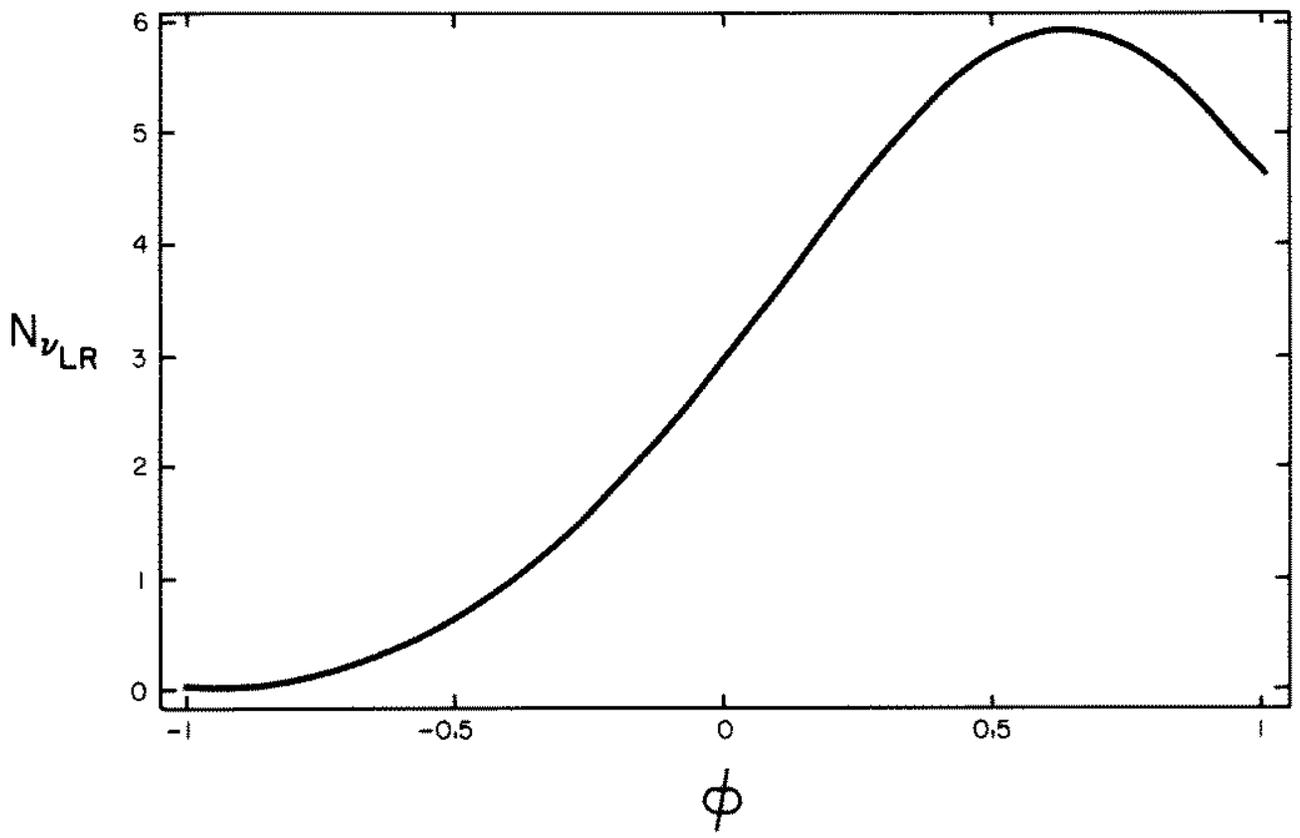
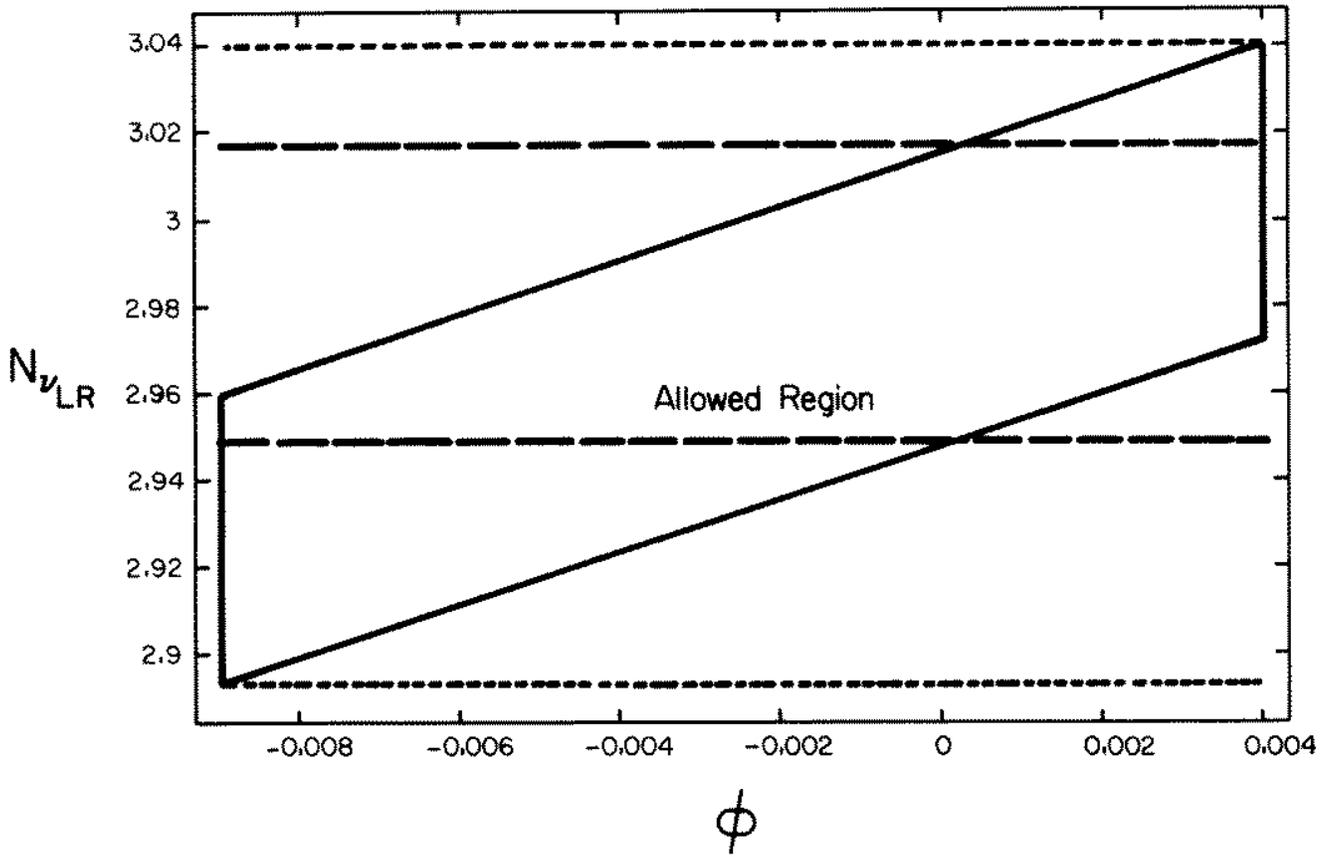
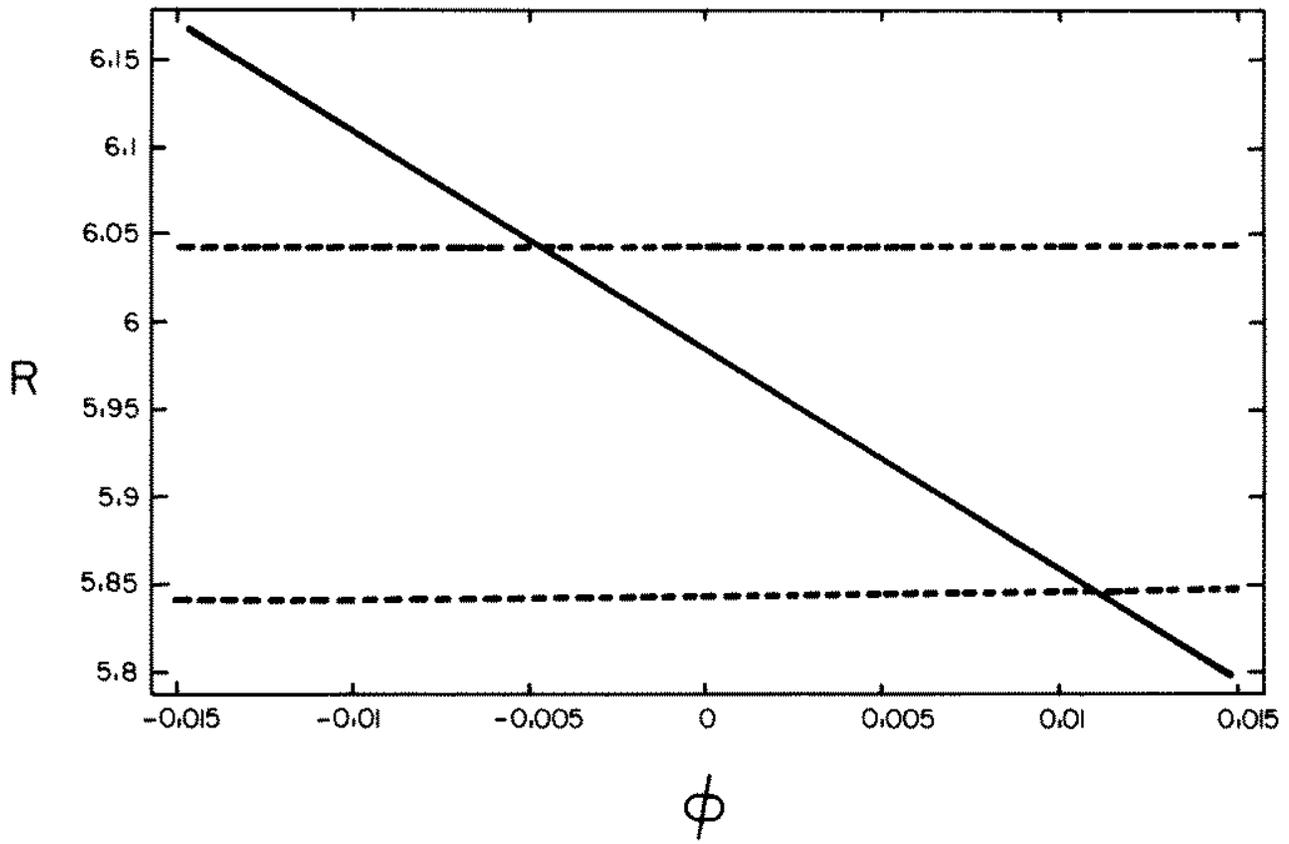
\end{document}